\documentclass[%
 reprint,
 amsmath,amssymb,
 aps,
]{revtex4-2}

\usepackage{graphicx}
\usepackage{dcolumn}
\usepackage{bm}
\usepackage[dvipsnames,table]{xcolor} %
\usepackage{physics}
\usepackage{braket}
\usepackage[breaklinks, %
colorlinks, linkcolor=Blue, citecolor=Blue, urlcolor=Blue]{hyperref} %
\usepackage{tabularx} %
\usepackage{booktabs} %
\usepackage{float} %
\newcommand{\bea}{\begin{eqnarray}}
\newcommand{\eea}{\end{eqnarray}}
\newcommand{\eq}[1]{Eq.~(\ref{#1})} %

\newcommand{\CR}[1]{\hat a^{\dagger}_{#1}}
\newcommand{\AN}[1]{\hat a_{#1}}
\newcommand{\eps}{\epsilon}

\begin{document}
\setcitestyle{super} 

\preprint{APS/123-QED}

\title{Extension of exactly-solvable Hamiltonians using symmetries of Lie algebras}

\author{Smik Patel}
\affiliation{Chemical Physics Theory Group, Department of Chemistry, University of Toronto, Toronto, Ontario M5S 3H6, Canada}
\affiliation{Department of Physical and Environmental Sciences, University of Toronto Scarborough, Toronto, Ontario M1C 1A4, Canada}
\author{Tzu-Ching Yen}
\affiliation{Chemical Physics Theory Group, Department of Chemistry, University of Toronto, Toronto, Ontario M5S 3H6, Canada}
\affiliation{Department of Physical and Environmental Sciences, University of Toronto Scarborough, Toronto, Ontario M1C 1A4, Canada}
\author{Artur F. Izmaylov}
\affiliation{Chemical Physics Theory Group, Department of Chemistry, University of Toronto, Toronto, Ontario M5S 3H6, Canada}
\affiliation{Department of Physical and Environmental Sciences, University of Toronto Scarborough, Toronto, Ontario M1C 1A4, Canada}

\date{\today}

\begin{abstract}
Exactly-solvable Hamiltonians that can be diagonalized using relatively simple unitary transformations are of great use in quantum computing. They can be employed for decomposition of interacting Hamiltonians either in Trotter-Suzuki approximations of the evolution operator for the quantum phase estimation algorithm, or in the quantum measurement problem for the variational quantum eigensolver. One of the typical forms of exactly solvable Hamiltonians is a linear combination of operators forming a modest size Lie algebra. Very frequently such linear combinations represent non-interacting Hamiltonians and thus are of limited interest for describing interacting cases. Here we propose the extension where coefficients in these combinations are substituted by polynomials of the Lie algebra symmetries. This substitution results in a more general class of solvable Hamiltonians and for qubit algebras is related to the recently proposed non-contextual Pauli Hamiltonians. In fermionic problems, this substitution leads to Hamiltonians with eigenstates that are single Slater determinants but with different sets of single-particle states for different eigenstates. The new class of solvable Hamiltonians can be measured efficiently using quantum circuits with gates that depend on the result of a mid-circuit measurement of the symmetries.
\end{abstract}

\maketitle


\section{Introduction}

Exactly solvable Hamiltonians are useful in the development of algorithms for approximately solving the Schr\"{o}dinger equation on both classical and quantum computers. Exact solvability in this context means that the unitary which diagonalizes the Hamiltonian can be obtained efficiently, either analytically or via an efficient numerical approach. In classical computing, solutions of exactly solvable Hamiltonians can serve as first order approximations which can be improved via perturbation theory. In quantum computing, a decomposition of the target Hamiltonian $\hat{H} = \sum_\alpha \hat{H}_\alpha$ into a sum of exactly solvable fragments has been used previously for computation of $\langle \hat H\rangle$ via quantum measurements\cite{izmaylovUnitaryPartitioningApproach2020, zhaoMeasurementReductionVariational2020,Verteletskyi:2020do, Yen2019b, huggins2019efficient, yenCartanSubalgebraApproach2021, Crawford2021efficientquantum, JenaFC, Choi2022ghostCS, Yen2023Deterministic, Choi2023fluidfermionic} in the variational quantum eigensolver \cite{peruzzoVariationalEigenvalueSolver2014} (VQE), and the Trotter\cite{lloydUniversalQuantumSimulators1996, martinez-martinezAssessmentVariousHamiltonian2022} and linear combinations of unitaries (LCU)\cite{childsHamiltonianSimulationUsing2012, loaizaReducingMolecularElectronic2023} decompositions of $e^{-i\hat{H}t}$ in quantum phase estimation\cite{kitaevQuantumMeasurementsAbelian1995, abramsQuantumAlgorithmProviding1999} (QPE). It has been shown that the choice of decomposition can greatly affect both the efficiency and the accuracy of the ground state estimation.\cite{Choi2022ghostCS,Choi2023fluidfermionic,martinez-martinezAssessmentVariousHamiltonian2022,loaizaReducingMolecularElectronic2023} This motivates the search for Hamiltonians that are exactly solvable by relatively simple unitary transformations. Discovery of new classes of such Hamiltonians will allow for further improvements in the efficiency and accuracy of quantum simulation algorithms.    

To keep the following discussion concrete, we will focus on decompositions of fermionic electronic structure Hamiltonians, although generalizations to other many-body problems are possible. In the second quantized formalism, electronic Hamiltonians are written as polynomials of the fermionic creation and annihilation operators: 
\begin{equation}
    \hat{H}_e = \sum_{pq} h_{pq} \hat{a}_p^\dagger \hat{a}_q + \sum_{pqrs} g_{pqrs} \hat{a}_p^\dagger \hat{a}_q \hat{a}_r^\dagger \hat{a}_s,
\end{equation}
where $h_{pq}$ and $g_{pqrs}$ are one- and two-electron integrals, and $p,q,r,s$ are indices enumerating the $N$ spin-orbitals. In quantum computing applications, $\hat{H}_e$ can be written as a linear combination of Pauli operators on $N$ qubits via a fermion-to-qubit map like the Jordan-Wigner\cite{jordanUeberPaulischeAequivalenzverbot1928} or Bravyi-Kitaev\cite{Bravyi_BKTransf,seeleyBravyiKitaevTransformationQuantum2012} transformations. Exactly solvable fragments of $\hat{H}_e$ can be constructed by considering the unitaries that can solve them. Two classes of unitaries have been useful: (1) Clifford group unitaries, (2) elements of Lie groups that have poly($N$) independent generators. In what follows we review several methods of obtaining exactly solvable Hamiltonians for partitioning of electronic Hamiltonians. The purpose of this exposition is to introduce ideas that will be used for a new class of exactly solvable Hamiltonians. 

In the fermionic algebra, one-electron Hamiltonians, also called free-fermionic Hamiltonians, can be solved using Lie groups that have a dimension that is polynomial in the system size:
\bea
\hat{h}_{1e} &=& \sum_{p,q=1}^{N} \tilde{h}_{pq} \CR{p}\AN{q} = \hat{U}^\dagger \left( \sum_{p=1}^{N} \eps_p \CR{p}\AN{p} \right)\hat{U},\\ \label{eq:U1e}
\hat{U} &=& \prod_{p>q} e^{\theta_{pq}[\CR{p}\AN{q} - \CR{q}\AN{p}]},
\eea  
where $\hat{U}$ is an orbital rotation from the Spin Lie group and the number operators $\CR{p}\AN{p}=\hat{n}_p$ are diagonal in the occupation number basis of the Fock space. $\hat{h}_{1e}$ is an element of the complexificaton of a compact $u(N)$ Lie algebra with anti-Hermitian generators $\{\CR{p}\AN{q}-\CR{q}\AN{p}\}$ and $\{i(\CR{p}\AN{q}+\CR{q}\AN{p})\}$. The associated Cartan subalgebra is generated by the number operators $\{i\hat{n}_p\}$. Exact solvability then follows from the maximal torus theorem, which implies that such Hamiltonians can be brought to a linear combination of Cartan subalgebra generators by conjugation via an element in the corresponding Lie group. In this case, due to compactness of the Lie algebra and the assumption that $\tilde{h}_{pq}$ is real, only the anti-symmetric generators $\CR{p}\AN{q} - \CR{q}\AN{p}$ are necessary to construct $\hat{U}$. 

The unitary $\hat U$ in \eq{eq:U1e} can be used to solve two-electron Hamiltonians that have the following special form\cite{yenCartanSubalgebraApproach2021} 
\bea
\hat{h}_{2e} &=& \hat{U}^\dagger \left(\sum_{pq} \lambda_{pq} \hat{n}_p \hat{n}_q \right) \hat{U} \label{factored}\\
&=& \sum_{pq} \tilde{h}_{pq} \CR{p}\AN{q} + \sum_{pqrs} \tilde{g}_{pqrs} \CR{p}\AN{q}\CR{r}\AN{s}. \label{expanded} 
\eea
Equation~\ref{expanded} has the form of a general two-electron Hamiltonian, but solvability of $\hat{h}_{2e}$  additionally constrains the constants $\tilde{h}_{pq}$ and $\tilde{g}_{pqrs}$. To demonstrate that the two-electron Hamiltonian $\hat{h}_{2e}$ is exactly-solvable using orbital rotations, it is sufficient to find $N$ one-electron symmetry operators $\{\hat{s}_p\}$ which commute with $\hat{h}_{2e}$ and are unitarily equivalent to occupation number operators, $\hat{s}_p = \hat{U}^\dagger \hat{n}_p \hat{U}$. This allows us to write $\hat{h}_{2e} = \sum_{pq} \lambda_{pq} \hat{s}_p \hat{s}_q$ as a quadratic polynomial of the symmetries.

In the qubit space, a formal analogue of one-electron unitary rotations are single-qubit rotations
\bea
\hat{U}_{1q} = \prod_{p=1}^{N} e^{i\theta_p(\bar{\sigma}_p,\bar{\tau}_p)}, 
\eea
where $\bar{\sigma}_p = (\hat{x}_p,\hat{y}_p,\hat{z}_p)$ is a vector of Pauli operators for the $p^{\rm th}$ qubit and $\bar{\tau}_p = (\tau_{xp},\tau_{yp},\tau_{zp})$ is a unit vector. However, the qubit Hamiltonians that are exactly solvable by single qubit rotations are not practical to work with. Such Hamiltonians, which are transformed to a polynomial of $\hat{z}_p$ operators by a single-qubit rotation, have exponentially many Pauli terms in general, and thus cannot be manipulated on a classical computer without additional constraints.

An alternative is to consider unitaries from the Clifford group. Such unitaries transform a single Pauli operator into a single Pauli operator and are more useful in defining exactly solvable $N$-qubit Hamiltonians. The associated exactly solvable Hamiltonians are linear combinations of mutually commuting Pauli operators, since it is possible to find a Clifford unitary that transforms such a linear combination into a polynomial function of $\hat{z}_p$ operators.\cite{Yen2019b,JenaFC} Thus, one can search for such exactly solvable fragments in the Hamiltonian of interest and measure or exponentiate them efficiently after applying the associated Clifford unitary. Like in the fermionic case, we can characterize this exact solvability in terms of symmetries, as any linear combination of mutually commuting Pauli operators $\hat{H}_q$ has $N$ algebraically independent Pauli symmetry operators $\hat{s}_p^{(q)}$ that can all be transformed to $\hat{z}_p$ using the same Clifford unitary $\hat{V}_c$: $\hat{s}_p^{(q)} = \hat{V}_c^\dagger \hat{z}_p \hat{V}_c$. Then, one can write $\hat{H}_q = p(\hat{s}_1^{(q)},\ldots,\hat{s}_N^{(q)})$ as a degree $N$ polynomial of the symmetry operators, which, if obtained as a measurable fragment of $\hat{H}_e$, will not have more than polynomial in $N$ terms. 

Another useful class of exactly solvable qubit Hamiltonians are those formed from linear combinations of anti-commuting Pauli operators.\cite{izmaylovUnitaryPartitioningApproach2020,zhaoMeasurementReductionVariational2020,loaizaReducingMolecularElectronic2023} Such Hamiltonians can be found as fragments using the anti-commutativity relation between Pauli operators and solved using elements of a small Lie group. Any Hamiltonian $\hat{H}_A = \sum_{i=1}^{L} c_i \hat{A}_i$ with real coefficients constructed from a set of mutually anti-commuting Pauli operators $\{\hat{A}_i\}$ can be rotated to a single Pauli operator present in the linear combination (e.g. $\hat{A}_L$) up to a scaling factor $||\vec{c}|| = \sqrt{\sum_{i=1}^{L} c_i^2}$. Solvability of $\hat{H}_A$ via a Lie group unitary is a consequence of the mutually anti-commuting Pauli operators being a part of the $so(L+1)$ Lie algebra generated by $\{\hat{A}_i\} \cup \{\hat{A}_i \hat{A}_j\}_{i > j}$. A downside to this grouping is that the maximum number of mutually anti-commuting Pauli operators acting on $N$ qubits is $2N+1$,\cite{bonet-monroigNearlyOptimalMeasurement2020} while the same number for mutually commuting Pauli operators is $2^N$. Therefore, usually the commuting Pauli operator groups are larger if used for a partitioning of a particular Hamiltonian. 

Recently, yet another class of exactly solvable Hamiltonians called non-contextual Pauli Hamiltonians was suggested, which are a linear combination of Pauli operators subject to some constraints defined in terms of commutation and anti-commutation relations between the Pauli terms (i.e. non-contextuality condition).\cite{Kirby2019_contextual_test, Kirby2020_contextual_simulation, Kirby2021_contextual_subspace, Weaving2022_contextual_stabilizer, Ralli2022contextual_unitary_partitioning}  The name ``non-contextual'' arises from the connection between properties of Pauli terms in these Hamiltonians and notions related to quantum contextuality and hidden variables theories.\cite{bellEinsteinPodolskyRosen1964, bellProblemHiddenVariables1966, peresTwoSimpleProofs1991, merminSimpleUnifiedForm1990, merminHiddenVariablesTwo1993} Even though these are interesting connections, they are not relevant to the findings of this work. Non-contextual Pauli Hamiltonians can be seen as generalizations of the Hamiltonians obtained from the grouping of mutually commuting or anti-commuting Pauli products. This motivates the question of determining algebraic reasons for exact solvability of non-contextual Pauli Hamiltonians. In this work, we present a framework that explains exact solvability of non-contextual Pauli Hamiltonians and provides more general conditions for solvability of qubit and fermionic Hamiltonians. 

The rest of this paper is organized as follows. In Sec.~\ref{theory_sec}, we construct our new class of exactly solvable Hamiltonians starting from the existence of an abelian symmetry group for a Lie algebra. In Sec.~\ref{app_q}, we demonstrate how non-contextual Hamiltonians arise from the application of this formalism to the $so(L+1)$ algebra associated to a set of anti-commuting Pauli operators. In Sec.~\ref{app_f}, we apply this formalism to construct fermionic exactly solvable Hamiltonians whose eigenstates are Slater determinants. In Sec.~\ref{app_circ}, we present quantum circuits for diagonalizing and measuring our exactly solvable Hamiltonians. In Sec.~\ref{res}, we present numerical results for decomposition of molecular electronic Hamiltonians into exactly solvable fragments. In Sec. \ref{con}, we conclude and address further directions to explore. 

\section{\label{theory_sec}Theory}

Our construction of new exactly solvable Hamiltonians is based on combination of known exactly solvable parts $\hat{H}_i$ into a larger Hamiltonian. This combination is based on the following result. Let $\hat{H}_i$ denote a set of Hamiltonians which are diagonalized by $\hat{U}_i$: $\hat{H}_i = \hat{U}_i^\dagger \hat{D}_i \hat{U}_i$, where $\hat{U}_i = e^{\hat{X}_i}$, and $\hat{X}_i$ is anti-Hermitian. Let $\hat{P}_i$ denote a complete set of orthogonal projectors such that for all $i,j$, $[\hat{H}_i, \hat{P}_j] = 0$. Then the Hamiltonian
\begin{equation}
    \hat{H} = \sum_i \hat{H}_i \hat{P}_i \label{proj-ham}
\end{equation}
is diagonalized by 
\begin{equation}
    \hat{U} = \sum_i \hat{U}_i \hat{P}_i = e^{\sum_i \hat{X}_i \hat{P}_i},
\end{equation}
and the corresponding diagonal form is
\begin{equation}
    \hat{U}\hat{H}\hat{U}^\dagger = \sum_i \hat{D}_i \hat{P}_i. \label{proj_solve}
\end{equation}
The proofs of these statements are given in Appendix \ref{appendix_a}.

For application of this result to the construction of exactly solvable Hamiltonians, 
note that any finite dimensional quantum many-body Hamiltonian can be expressed as a 
polynomial of Hermitian operators $\{\hat{\Lambda}_i\}$ on a Hilbert space $V$ which generate a 
compact Lie algebra $\mathcal{A}$. For compact Lie algebras, we can introduce the Cartan subalgebra $\mathcal{C} \subset \mathcal{A}$, which is any maximal abelian Lie subalgebra of $\mathcal{A}$. Since $\mathcal{C}$ is abelian, polynomials of generators of $\mathcal{C}$ can in principle be measured simultaneously. A class of exactly solvable Hamiltonians are the Hamiltonians which are themselves elements of $\mathcal{A}$ up to a constant shift, implying that they are expressible as:
\begin{equation}
    \hat{H}_{\vec{c}} = \sum_i c_i \hat{\Lambda}_i + c_0,
\end{equation}
The addition of a constant shift allows for additional flexibility while not affecting the overall exact solvability. Exact solvability of $\hat{H}_{\vec{c}}$ follows from the maximal torus theorem for compact groups, which implies that any $\hat{H}_{\vec{c}}$ can be transformed to an element of the Cartan subalgebra up to the constant shift by a unitary in the corresponding Lie group.\cite{hallLieGroupsLie2015} Application of this formalism to quantum computing requires transformations which map the Cartan subalgebra eigenbasis to the quantum computer's computational basis. Such transformations are known for algebras considered in this work. 

Our extension of the exact solvability obtained from the maximal torus theorem takes into account the representation defined by $V$ of the Lie algebra $\mathcal{A}$. If the representation is irreducible, meaning that there are no non-trivial invariant subspaces of $V$ fixed by all $\hat{X} \in \mathcal{A}$, then Schur's lemma implies that the only operators that commute with all $\hat{X}$ are scalar multiples of the identity on $V$. In the case that the representation is reducible, we can decompose $V$ into a direct sum of invariant subspaces $V = \oplus_\lambda V_\lambda$ which are fixed by all $\hat{X}$. Then, any operator $\hat{S} = \oplus_\lambda c_\lambda \hat{I}_\lambda$, for $c_\lambda \in \mathbb{C}$, which acts as a scalar multiple of the identity $\hat{I}_\lambda$ on all the invariant subspaces will commute with all elements of $\mathcal{A}$, implying that $\mathcal{A}$ has non-trivial symmetries. Thus, when the representation is reducible, we can introduce a non-trivial abelian symmetry group of Hermitian operators $\mathcal{G}$ with minimal generating set $G = \{\hat{C}_1,\ldots,\hat{C}_K\}$ such that the $\hat{C}_k$ commute with the $\hat{A}_i$. $\mathcal{G}$ is finite precisely when the $\hat{C}_k$ are reflection operators (i.e. have spectrum contained in $\{1,-1\}$), since $\{1, -1\}$ is the only finite subgroup of $\mathbb{R} \setminus \{0\}$ under multiplication. In this case, $\mathcal{G}$ is a representation of the group $\oplus_{l=1}^{K} \mathbb{Z}_2$ on $V$. For the algebras considered in this work, all symmetry group elements are unitarily equivalent to Pauli operators which satisfy this constraint. The elements of $G$ can be simultaneously spectral decomposed
\begin{equation}
    \hat{C}_k = \sum_{\vec{v}} v_k \hat{P}_{\vec{v}},
\end{equation}
where $\vec{v}$ denotes an assignment of an eigenvalue $v_k$ to all $\hat{C}_k$ simultaneously, and $\hat{P}_{\vec{v}}$ is the projection operator onto the intersection of all the corresponding eigenspaces, herein referred to as a ``simultaneous eigenspace''. Each simultaneous eigenspace defines an invariant subspace of $V$ for $\mathcal{A}$, but are not necessarily irreducible. Any eigenbasis of $G$ is a valid eigenbasis for all operators in the group algebra $\mathbb{R}[\mathcal{G}]$, defined to be the space of real linear combinations of elements of $\mathcal{G}$. The reason is, any operator in $\mathbb{R}[\mathcal{G}]$ can be expressed as a polynomial of the group generators
\begin{equation}
    p(\hat{C}_1,\ldots,\hat{C}_K) = \sum_l d_l \hat{C}_l + \sum_{lm} d_{lm} \hat{C}_l \hat{C}_m + \cdots
\end{equation}
implying that $p(\hat{C}_1,\ldots,\hat{C}_K)$ admits the following spectral decomposition
\begin{equation}
    p(\hat{C}_1,\ldots,\hat{C}_K) = \sum_{\vec{v}} p(v_1,\ldots,v_K) \hat{P}_{\vec{v}} = \sum_{\vec{v}} p(\vec{v}) \hat{P}_{\vec{v}}.
\end{equation}
When $\mathcal{G}$ is infinite, this polynomial can have arbitrarily many terms in general, but can always be brought to not more than $\dim(V)$ terms by identifying a maximal set of linearly independent group elements, and expressing all other group elements present in $p(\hat{C}_1,\ldots,\hat{C}_K)$ in terms of this independent set. The existence of the abelian symmetry group $\mathcal{G}$ for $\mathcal{A}$ allows us to define a class of exactly solvable Hamiltonians by starting with the general form $\hat{H}_{\vec{c}}$ of Hamiltonians solvable via the maximal torus theorem, but replacing the real number coefficients $c_0, c_i$ by elements of $\mathbb{R}[\mathcal{G}]$:
\begin{equation}
\hat{H}_{\vec{p}} = \sum_i p_i(\hat{C}_1,\ldots,\hat{C}_K)\hat{\Lambda}_i + p_0(\hat{C}_1,\ldots,\hat{C}_K). \label{esm}
\end{equation}
The abelian group generators $\hat{C}_k$ thus form a set of $K$ algebraically independent symmetries for the Hamiltonian $H_{\vec{p}}$, and, when working within a simultaneous eigenspace in which all the symmetries act as eigenvalues, the resulting operator is solvable via a Lie group rotation. More formally, exact solvability of $H_{\vec{p}}$ follows from the result in \eq{proj_solve}. Substitution of the spectral decompositions of the elements of $\mathbb{R}[\mathcal{G}]$ into the Hamiltonian $\hat{H}_{\vec{p}}$ yields:
\begin{equation}
    \hat{H}_{\vec{p}} = \sum_{\vec{v}}\left[\sum_i p_i(\vec{v})\hat{\Lambda}_i + p_0(\vec{v})\right]\hat{P}_{\vec{v}} = \sum_{\vec{v}} \hat{H}_{\vec{v}} \hat{P}_{\vec{v}},
\end{equation}
where the $\hat{H}_{\vec{v}}$ are elements of $\mathcal{A}$ up to a constant shift. Thus, they are solvable by a Lie group element $\hat{U}_{\vec{v}}$ generated by $\hat{X}_{\vec{v}}$. Then, the commutation between elements of $\mathcal{A}$ and elements of $\mathcal{G}$ implies, by \eq{proj_solve}, that $\hat{H}_{\vec{p}}$ is transformed into a polynomial of the Cartan subalgebra $\mathcal{C}$ generators and the group generators $G$ via $\hat{V} = \exp\big(\sum_{\vec{v}} \hat{X}_{\vec{v}} \hat{P}_{\vec{v}}\big)$.

\section{Applications}\label{app_sec}

In this section, we illustrate how this formalism can be used to construct new measurable many-body operators for quantum computing applications in both the qubit algebra, generated by Pauli operators, and the fermionic algebra, generated by single-particle fermionic excitation operators. 

\subsection{Qubit Algebra}\label{app_q}

Introduction of Pauli symmetries of the $so(L+1)$ Lie algebra generated by mutually anti-commuting Pauli operators $\{\hat{A}_i\}_{i=1}^{L}$ allows us to construct non-contextual Pauli Hamiltonians. A set $\mathcal{S}$ of Pauli operators is non-contextual if it decomposes as $\mathcal{S} = \mathcal{Z} \cup \mathcal{T}$, where elements of $\mathcal{Z}$ commute with elements of $\mathcal{S}$, and commutation is an equivalence relation on $\mathcal{T}$, implying that $\mathcal{T}$ decomposes into disjoint commutation classes $\mathcal{C}_1 \cup \cdots \cup \mathcal{C}_N$, where operators in the same class commute, and operators in distinct classes anti-commute. A Pauli Hamiltonian is called non-contextual if its Pauli terms come from a non-contextual set. To construct non-contextual Pauli Hamiltonians using symmetries of the anti-commuting set, let $\mathcal{G}$ denote an abelian group of Pauli operators with some minimal generating set $G = \{\hat{C}_1,\ldots,\hat{C}_K\}$ whose elements also commute with the $\hat{A}_i$. Using elements of the group algebra $\mathbb{R}[\mathcal{G}]$ as coefficients of the anti-commuting Pauli operators yields Pauli Hamiltonians of the following form:
\begin{equation}
    \hat{H} = \sum_{i=1}^{L} p_i(\hat{C}_1,\ldots,\hat{C}_K)\hat{A}_i + p_0(\hat{C}_1,\ldots,\hat{C}_K) \label{qubit_nc}.
\end{equation}
To prove that $\hat{H}$ is a non-contextual Pauli Hamiltonian, let $\mathcal{S}$ denote the set of Pauli operators present in \eq{qubit_nc}. Any Pauli operator present in $\hat{H}$ is either an element of $\mathcal{G}$ or one of the disjoint cosets $\mathcal{G}\hat{A}_i = \{\hat{X}\hat{A}_i : \hat{X} \in \mathcal{G}\}$, implying that we can write $\mathcal{S} = \mathcal{Z} \cup \mathcal{T}$, where $\mathcal{Z} \subset \mathcal{G}$ contains operators which commute with all operators in $\mathcal{S}$, and $\mathcal{T} \subset \mathcal{G}\hat{A}_1 \cup \cdots \cup \mathcal{G}\hat{A}_L$ is a union of disjoint commutation classes. Since Pauli operators either commute or anti-commute, it follows that for $i \not= j$, operators from $\mathcal{G}\hat{A}_i$ anti-commute with all operators in $\mathcal{G}\hat{A}_j$. Thus, $\mathcal{S}$ is non-contextual, implying that $\hat{H}$, whose Pauli terms come from $\mathcal{S}$, is a non-contextual Pauli Hamiltonian. Moreover, as shown in \citet{Kirby2020_contextual_simulation}, any non-contextual Pauli Hamiltonian can be brought to this form (see appendix \ref{appendix_b} for the derivation). Thus, this construction characterizes the non-contextual Pauli Hamiltonians. To diagonalize $\hat{H}$, first, apply the Clifford group unitary $\hat{U}_\text{T}$ that transforms $\hat{C}_1, \ldots, \hat{C}_K$ to Pauli-$\hat{z}$ operators on the first $K$ qubits, and the anti-commuting set $\{\hat{A}_i\}$ to Pauli operators $\{\hat{A}'_i\}$ on the remaining qubits:
\begin{equation}
    \hat{U}_\text{T} \hat{H} \hat{U}_\text{T}^\dagger = \sum_{i=1}^{L} p_i(\hat{z}_1,\ldots,\hat{z}_K) \hat{A}'_i + p_0(\hat{z}_1,\ldots,\hat{z}_K).
\end{equation}
$\hat{U}_\text{T}$ can be found using techniques developed for qubit tapering.\cite{bravyiTaperingQubitsSimulate2017a} The new group generators $\hat{z}_1,\ldots,\hat{z}_K$ can be simultaneously spectral decomposed as follows:
\begin{equation}
    \hat{z}_k = \sum_{\vec{v}} v_k \hat{P}_{\vec{v}}, \qquad \hat{P}_{\vec{v}} = \prod_{l=1}^{K} \frac{1 + v_l\hat{z}_l}{2},
\end{equation}
for $\vec{v} \in \{-1,1\}^K$. Then, the non-contextual Pauli Hamiltonian can be written:
\begin{equation}
    \hat{U}_\text{T} \hat{H} \hat{U}_\text{T}^\dagger = \sum_{\vec{v}} \left[\sum_{i=1}^{L} p_i(\vec{v})\hat{A}'_i + p_0(\vec{v})\right]\hat{P}_{\vec{v}} = \sum_{\vec{v}} \hat{H}_{\vec{v}} \hat{P}_{\vec{v}}.
\end{equation}
The operators $\hat{H}_{\vec{v}}$ are linear combinations of anti-commuting Pauli operators up to a constant shift $p_0(\vec{v})$. The unitary $\hat{U}_{\vec{v}}$ that rotates $\hat{H}_{\vec{v}}$ to a single Pauli operator $\hat{A}_L'$ up to $p_0(\vec{v})$ and the normalization constant $a(\vec{v}) = \sqrt{\sum_{i=1}^{L} p_i^2(\vec{v})}$ has the following form
\begin{equation}
    \hat{U}_{\vec{v}} = \prod_{j=1}^{L - 1} e^{\theta_j(\vec{v}) \hat{A}_L' \hat{A}_j' / 2} \label{so_solver},
\end{equation}
where the angles $\theta_j(\vec{v})$ are obtained from the $p_i(\vec{v})$ via the following equations:
\begin{align}
    p_1(\vec{v}) \cos(\theta_1(\vec{v})) &= p_L(\vec{v}) \sin(\theta_1(\vec{v}))\nonumber\\
    p_j(\vec{v}) \cos(\theta_j(\vec{v})) &= \sin(\theta_j(\vec{v}))\sqrt{\sum_{i=1}^{j-1}p_i^2(\vec{v}) + p_L^2(\vec{v})}, \ j > 1.
\end{align}
An additional single qubit rotation $\hat{R}$ that rotates $\hat{A}_L'$ to a product $\hat{z}_S = \prod_{s \in S} \hat{z}_s$ of Pauli-$\hat{z}$ operators on a subset of qubits $S$ will solve $\hat{H}_{\vec{v}}$. Thus
\begin{align}
    \hat{R} \hat{U}_{\vec{v}} \hat{H}_{\vec{v}} \hat{U}_{\vec{v}}^\dagger \hat{R}^\dagger &= a(\vec{v}) \hat{z}_S + p_0(\vec{v}).
\end{align}
Since \eq{so_solver} is given in a product form, we can use \eq{prod_form} to write:
\begin{equation}
    \hat{V} = \sum_{\vec{v}} \hat{U}_{\vec{v}} \hat{P}_{\vec{v}} = \prod_{\vec{v}} \prod_{j=1}^{L - 1} e^{\theta_j(\vec{v}) \hat{A}_L' \hat{A}_j' \hat{P}_{\vec{v}} / 2},
\end{equation}
from which we arrive at:
\begin{equation}
    \hat{R}\hat{V}\hat{U}_\text{T} \hat{H} \hat{U}_\text{T}^\dagger \hat{V}^\dagger \hat{R}^\dagger = \sum_{\vec{v}} \left[a(\vec{v}) \hat{z}_S + p_0(\vec{v})\right] \hat{P}_{\vec{v}}.
\end{equation}

\subsection{Fermionic Algebra}\label{app_f}

A fermionic Hamiltonian which has an eigenbasis containing only Slater determinants is called mean-field solvable.\cite{izmaylovHowDefineQuantum2021} The formalism of this work gives us a compact representation of two-body fermionic mean-field solvable Hamiltonians, as well as a procedure to solve them. Start with a set $S = \{1,\ldots,N\}$ of $N$ spin orbitals, and partition them into two disjoint subsets $S_1, S_2$ of size $N - K$ and $K$. Let $i,j$ denote indices over $S_1$ and $k,l$ denote indices over $S_2$. One-body Hermitian operators on $S_1$ are exactly solvable:
\begin{equation}
    \hat{H} = \sum_{ij} h_{ij} \hat{a}_i^\dagger \hat{a}_j + h_0, \qquad h_{ij} = h_{ji}.
\end{equation}
$\hat{H}$ can be solved via an orbital rotation $\hat{U}_{\vec{\theta}} = e^{\hat{X}_{\vec{\theta}}}$ on $S_1$, which is defined by its anti-Hermitian generator:
\begin{equation}
    \hat{X}_{\vec{\theta}} = \sum_{i > j} \theta_{ij}(\hat{a}_i^\dagger \hat{a}_j - \hat{a}_j^\dagger \hat{a}_i).
\end{equation}
The angles $\theta_{ij}$ can be obtained from the matrix elements of the generator of the unitary matrix that diagonalizes the one-body-tensor $h_{ij}$. Then:
\begin{equation}
    \hat{U}_{\vec{\theta}} \hat{H} \hat{U}_{\vec{\theta}}^\dagger = \sum_i \gamma_i \hat{n}_i + h_0,
\end{equation}
where $\gamma_i$ are the eigenvalues of $h_{ij}$. Number operators $\hat{n}_k$ on $S_2$ form a set of independent symmetries of $\hat{H}$, and so, replacement of $h_{ij}$ and $h_0$ by polynomials of the $\hat{n}_k$ yield the following exactly solvable twobody Hamiltonian:
\begin{equation}
    \hat{H} = \hat{U}_\text{T}^\dagger \left[\sum_{ij} \left(\sum_k h_{ij}^{(k)} \hat{n}_k\right) \hat{a}_i^\dagger \hat{a}_j + \sum_{kl} \lambda_{kl} \hat{n}_k \hat{n}_l\right] \hat{U}_\text{T}, \label{ferm_mf}
\end{equation}
where $\hat{U}_\text{T}$ is an orbital rotation on all $N$ orbitals, and plays the same role in the fermionic case as the Clifford group rotation in the qubit case. Although the non-invertible $\hat{n}_k$ cannot be group generators, they can be transformed to reflections that generate an abelian symmetry group via $\hat{n}_k \mapsto \hat{z}_k = 2\hat{n}_k - 1$, and the polynomials one can form from $\hat{n}_k$ and $\hat{z}_k$ coincide. Moreover, note that substitution of higher degree polynomials of the $\hat{n}_k$ will yield more general exactly solvable $n$-body Hamiltonians. Solving $\hat{H}$ relies on simultaneous spectral decomposition of the number operators $\hat{n}_k$
\begin{equation}
    \hat{n}_k = \sum_{\vec{v}} v_k \hat{P}_{\vec{v}}, \qquad \hat{P}_{\vec{v}} = \prod_{l \in S_2} \hat{n}_l^{v_l} (1 - \hat{n}_l)^{1 - v_l},
\end{equation}
where $\vec{v} \in \{0,1\}^K$. Substitution of these spectral decompositions and rearranging yields:
\begin{align}
    \hat{U}_\text{T} \hat{H} \hat{U}_\text{T}^\dagger &= \sum_{\vec{v}} \left[\sum_{ij} \gamma_{ij}(\vec{v}) \hat{a}_i^\dagger \hat{a}_j + \sum_{kl} \lambda_{kl} v_k v_l\right] \hat{P}_{\vec{v}}\nonumber\\
    &= \sum_{\vec{v}} \hat{H}_{\vec{v}} \hat{P}_{\vec{v}},
\end{align}
where $\gamma_{ij}(\vec{v}) = \sum_k h_{ij}^{(k)} v_k$. The operators $\hat{H}_{\vec{v}}$ are onebody Hermitian operators on $S_1$, implying they are solvable by an orbital rotation $\hat{U}_{\vec{v}} = e^{\hat{X}_{\vec{v}}}$ with appropriate choice of angles $\theta_{ij}(\vec{v})$:
\begin{equation}
    \hat{H}_{\vec{v}} = e^{-\hat{X}_{\vec{v}}}\left[\sum_i \gamma_i(\vec{v}) \hat{n}_i + \sum_{kl} \lambda_{kl} v_k v_l\right] e^{\hat{X}_{\vec{v}}}.
\end{equation}
Writing $\hat{V} = \exp\left(\sum_{\vec{v}} \hat{X}_{\vec{v}}\hat{P}_{\vec{v}}\right)$, we have
\begin{equation}
    \hat{V}\hat{U}_\text{T} \hat{H} \hat{U}_\text{T}^\dagger \hat{V}^\dagger = \sum_{\vec{v}} \left[\sum_i \gamma_i(\vec{v}) \hat{n}_i + \sum_{kl} \lambda_{kl} v_k v_l\right] \hat{P}_{\vec{v}}.
\end{equation}
To obtain an eigenstate of $\hat{H}$, start with an electronic configuration on the $S_2$ orbitals, which specifies a value $\vec{v}$. Associated to $\vec{v}$ are $2^{|S_1|}$ eigenstates of $\hat{H}$ obtained by considering all electronic configurations on the remaining $|S_1|$ orbitals obtained by diagonalizing $\hat{H}_{\vec{v}}$.

\subsection{Quantum Circuits for Measurement}\label{app_circ}

\begin{figure}[h]
    \centering
    \includegraphics[width=\columnwidth]{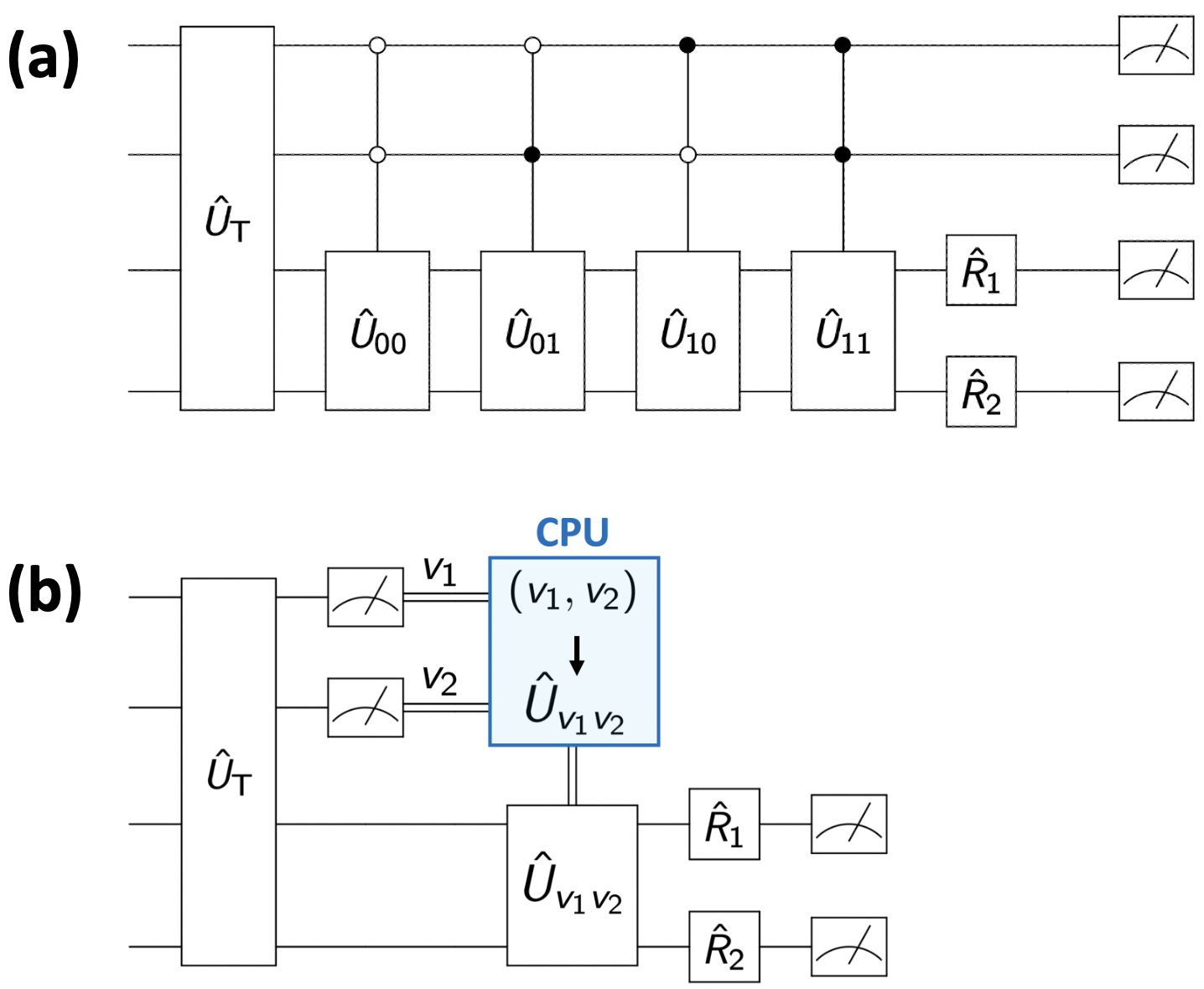}
    \caption{(a) Four-qubit example of the circuit for diagonalizing the exactly solvable Hamiltonians on a quantum computer. After application of $\hat{U}_\text{T}$, the fully commuting symmetries are transformed to $\hat{z}$'s on the top two qubits, and controlled unitaries for all computational basis states $\ket{\vec{v}}$ are used to diagonalize all the blocks $\hat{H}_{\vec{v}}$. In the fermionic case, $\hat{R}_1$ and $\hat{R}_2$ are the identity gate. (b) Dynamic quantum circuit for measuring exactly solvable Hamiltonian with a unitary $\hat{U}_{v_1 v_2}$ classically conditioned on a measurement result $(v_1, v_2)$ of the top two qubits. To obtain the circuit implementation of $\hat{U}_{v_1 v_2}$, a classical CPU must calculate the parameters of $\hat{U}_{v_1 v_2}$ from $(v_1, v_2)$.}
    \label{figure1}
\end{figure}

Application of this formalism to quantum computing requires quantum circuits for measuring these exactly solvable Hamiltonians. Considering both qubit and fermionic cases, this requires a quantum circuit decomposition for $\hat{V} = \exp\left(\sum_{\vec{v}} \hat{X}_{\vec{v}} \hat{P}_{\vec{v}}\right)$. To do this, use commutativity of the terms in the generator to write:
\begin{equation}
    \hat{V} = \prod_{\vec{v}} e^{\hat{X}_{\vec{v}} \hat{P}_{\vec{v}}}.
\end{equation}
Then, since every simultaneous eigenspace of $\hat{z}_1,\ldots,\hat{z}_K$ (or, equivalently, $\hat{n}_1,\ldots,\hat{n}_k$ via $\hat{z}_k = 2\hat{n}_k - 1$) is associated to exactly one computational basis state $\ket{\vec{v}}$ on the first $K$ qubits, we may write:
\begin{equation}
    \hat{P}_{\vec{v}} = \ket{\vec{v}}\bra{\vec{v}} \otimes 1.
\end{equation}
From \eq{basecase}, a straightforward calculation yields
\begin{equation}
    e^{\hat{X}_{\vec{v}} \hat{P}_{\vec{v}}} = \ket{\vec{v}}\bra{\vec{v}} \otimes \hat{U}_{\vec{v}} + \big(1 - \ket{\vec{v}}\bra{\vec{v}}\big),
\end{equation}
implying, from $\sum_{\vec{v}} \hat{P}_{\vec{v}} = 1$ and $P_{\vec{v}}P_{\vec{w}} = \delta_{\vec{v}\vec{w}}P_{\vec{w}}$, that $\hat{V} = \sum_{\vec{v}} \ket{\vec{v}}\bra{\vec{v}} \otimes \hat{U}_{\vec{v}}$. Thus, $e^{\hat{X}_{\vec{v}} \hat{P}_{\vec{v}}}$ is the unitary $\hat{U}_{\vec{v}}$ conditioned on the first $K$ qubits being in the state $\ket{\vec{v}}$. This means, to diagonalize non-contextual Pauli Hamiltonians, we need to implement controlled $e^{\theta_j(\vec{v}) \hat{A}_L' \hat{A}_j' / 2}$ rotations, and for fermionic mean-field solvable Hamiltonians, we need to implement controlled orbital rotations $e^{\hat{X}_{\vec{v}}}$. A schematic for the circuit that diagonalizes exactly solvable Hamiltonians is shown in Fig. \ref{figure1}a. Assuming no additional structure, implementation of this circuit is exponentially costly in the number of independent $\mathbb{Z}_2$ symmetries $K$. However, for application of these Hamiltonians to VQE, we only need to measure $\hat{H}$. For measuring $\hat{H}$, one can note that the qubits corresponding to the symmetries only play the role of control qubits for controlled unitaries after $\hat{U}_\text{T}$ is applied. Thus, the exponentially many controlled unitaries can be replaced by a single $\hat{U}_{\vec{v}}$ classically conditioned on the result $\vec{v}$ of a mid-circuit measurement of the independent symmetries, as shown in Fig. \ref{figure1}b. This requires a classical computation of the parameters for the remainder of the circuit which implements $\hat{U}_{\vec{v}}$ after the mid-circuit measurement of the symmetries. These so-called ``dynamic quantum circuits'' have been studied and implemented in superconducting\cite{corcolesExploitingDynamicQuantum2021, googlequantumaiExponentialSuppressionBit2021, acharyaSuppressingQuantumErrors2023} and trapped ion\cite{pinoDemonstrationTrappedionQuantum2021, mosesRaceTrackTrappedIon2023} quantum processors.

\section{Results}\label{res}

Here, we assess these new exactly solvable Hamiltonians by testing their capability to serve as measurable fragments of more complex Hamiltonians. A partitioning 
\bea
    \hat{H} = \sum_{\alpha=1}^{N_\text{frag}} \hat{H}_\alpha 
  \eea 
expresses the target Hamiltonian $\hat{H}$ as a sum of exactly solvable fragments $\hat{H}_\alpha$. By virtue of the exact solvability of the $\hat{H}_\alpha$, one can measure individual $\hat{H}_\alpha$ separately with a trial wavefunction $\ket{\psi}$ and thereby estimate the expectation value $\bra{\psi} \hat{H} \ket{\psi}$ as the sum of expectation values of the parts. Here we investigated the efficacy of using non-contextuality and fermionic mean-field solvability as the criterion for $\hat{H}_\alpha$ to partition electronic Hamiltonians. Details of the electronic Hamiltonians used in this work are presented in Appendix \ref{app_comp}.

For decompositions based on qubit algebra techniques, the Bravyi-Kitaev transformation was used to express $\hat{H}_e$ in terms of Pauli operators. For assessment of non-contextual fragments, we compared the number of fragments $N_\text{frag}$ obtained using the non-contextuality (NC) criterion to previously studied fully commutative\cite{JenaFC, Yen2019b} (FC) and anti-commutative\cite{zhaoMeasurementReductionVariational2020, izmaylovUnitaryPartitioningApproach2020} (AC) criterions. To obtain the fragments $\hat{H}_\alpha$, we used the greedy approach proposed by \citet{Crawford2021efficientquantum}. Table~\ref{tab:result_nfrag_qub} shows that aside from $\rm NH_3$, NC often reduces the number of fragments required to partition electronic Hamiltonians compared to FC. 
  
Note that FC and AC fragments are necessarily non-contextual, and FC outperforms AC. Therefore, one can  find a more efficient partitioning by adding Pauli operators to existing FC fragments if the addition preserves non-contextuality. Following this heuristic, we explored a partitioning (FNC) that exploits both the FC and NC criterions. The algorithm consists of three steps: (1) find the FC fragments in a greedy fashion, (2) sort the fragments in number of Pauli operators, and finally (3) add the Pauli operators of smaller fragments (containing less Pauli operators) to larger fragments (containing more Pauli operators) while preserving the non-contextuality criterion. Table~\ref{tab:result_nfrag_qub} shows that the FNC heuristics consistently outperforms FC by approximately $20\%$ in number of fragments required. 
  \begin{table}[htbp]
    \centering
    {\begin{tabularx}{\columnwidth}{@{\extracolsep{\fill}} c c c c c c}
        \toprule
         Systems & Pauli & AC & FC & NC & FNC \\
        \midrule
        H$_2$ & 15 & 11 & 2 & 1 & 1\\ 
        \midrule
         LiH & 631 & 110 & 42 & 34 & 31\\ 
        \midrule
        BeH$_2$& 666 & 134 & 36 & 33 & 29 \\
        \midrule
         H$_2$O & 1086 & 156 & 50 & 49 & 40 \\
        \midrule
         NH$_3$ & 3609 & 324 & 122 & 148 & 101 \\
        \bottomrule
        \end{tabularx}
    }
    \caption{Number of fragments found by greedy method with fully anti-commuting (AC), fully-commuting (FC), non-contextual (NC), and fully-commuting with non-contextual modification (FNC) fragments. The number of Pauli operators in each Hamiltonian is presented for comparison.}
    \label{tab:result_nfrag_qub}
  \end{table}

The advantage of NC can be seen not only for electronic Hamiltonians. Consider the following model Hamiltonian which is an extension of the Heisenberg spin Hamiltonian \cite{Kokcu_FixedDepthSimCartanDecomp}
  \bea
    \hat{H} &=& \sum_{i=1}^{2n-1} (
        a_i \hat{x}_i \hat{x}_{i+1} 
        + b_i \hat{y}_i \hat{y}_{i+1} 
        + c_i \hat{z}_i \hat{z}_{i+1} 
    ) \nonumber \\
    &+& \sum_{j=1}^{2n} d_j \hat{z}_j. 
  \eea 
Evidently, three FC fragments are required to partition $\hat{H}$: the first fragment contains terms in the first summation with odd $i$, the second fragment contains even $i$, and the third fragment contains simply $\sum_{j=1}^{2n} d_j \hat{z}_j$. In comparison, using non-contextual fragments, $\hat{H}$ can be written as the sum of only two solvable fragments: 
\bea
    \hat{H}^{(\text{NC)}}_1 &=&
    \sum_{i=1}^{n} (
        a_{2i-1} \hat{x}_{2i-1} \hat{x}_{2i} 
        + b_{2i-1} \hat{y}_{2i-1} \hat{y}_{2i}
     \nonumber \\
    &+& c_{2i-1} \hat{z}_{2i-1} \hat{z}_{2i}
        + d_{2i-1} \hat{z}_{2i-1} + d_{2i} \hat{z}_{2i}
    )
\\
    \hat{H}^{(\text{NC)}}_2 &=&
    \sum_{i=1}^{n} (
        a_{2i} \hat{x}_{2i} \hat{x}_{2i+1} 
        + b_{2i} \hat{y}_{2i} \hat{y}_{2i+1} 
     \nonumber \\
    &+& c_{2i} \hat{z}_{2i} \hat{z}_{2i+1}
    ) 
  \eea 
Note that $\hat{H}^{(\text{NC)}}_1$ is solvable since it is a sum of $n$ non-contextual Hamiltonians acting on disjoint sets of qubits. 
  
We also assessed the fermionic mean-field solvable Hamiltonians by comparing them to previously developed low-rank\cite{huggins2019efficient} (LR) and greedy full-rank\cite{yenCartanSubalgebraApproach2021} (GFRO) decompositions. We computed mean-field solvable fragments via a greedy approach, where each fragment was obtained via gradient minimization of the $L_2$ distance of the two-body-tensor $g_{pqrs}$ of $\hat{H}_e$ and the two-body-tensor of the fermionic mean-field fragment calculated from \eq{ferm_mf}. The decomposition was done until the squared $L_2$ norm of the remaining two body tensor was less than $10^{-6}$. The results are shown in table \ref{tab:result_nfrag_ferm}. To define the mean-field fragments, first, we need to specify a partitioning of $N$ spin orbitals $S$ into two disjoint subsets $S_1, S_2$ of size $N - K$ and $K$, which defines the GMF($K$) decomposition. We used an ordering of the orbitals from lowest energy to highest energy, an $\alpha\beta$ ordering of the spins, and we used the even partitioning of the spin orbitals $S_1 = \{1,\ldots, N/2\}, S_2 = \{N/2 + 1,\ldots,N\}$. We note here that the specific choice of partitioning of the spin orbitals up to choosing the size parameter $K$ did not greatly affect the results. The results suggest that the GFRO and LR criterions outperform the GMF criterion. The GFRO fragments are a special case of the mean-field solvable fragments where $K = N$, so that $S_1 = \varnothing$. Indeed, we found empirically that the GMF results consistently improve as $K$ gets larger, and converge identically to the GFRO results in the $K = N$ case. An explanation for this is to note that the majority of the $L_2$ norm of $g_{pqrs}$ for the systems considered comes from the $g_{ppqq}$ terms, and a single GFRO fragment is sufficient to capture the entirety of this contribution in the orbital basis where the contribution is greatest, whereas no other GMF fragment has this property, due to the disjointness of $S_1$ and $S_2$.

  \begin{table}[htbp]
    \centering
    {\begin{tabularx}{\columnwidth}{@{\extracolsep{\fill}} c c c c }
        \toprule
         Systems & GFRO & LR & GMF($N/2$) \\
        \midrule
        H$_2$
            & 3 & 4 & 15\\ 
        \midrule
         H$_4$
             & 15 & 8 & 20\\ 
        \midrule
        LiH
            & 24 & 20 & 58 \\
        \midrule
        BeH$_2$
            & 32 & 23 & 83 \\
        \midrule
        H$_2$O
            & 34 & 24 & 102 \\
        \midrule
        NH$_3$
            & 47 & 28 & 61 \\
        \bottomrule
        \end{tabularx}
    }
    \caption{Number of fragments found using greedy full-rank (GFRO), low-rank (LR) and greedy mean-field with even splitting of $N$ orbitals into two disjoint subsets (GMF($N/2$)).}
    \label{tab:result_nfrag_ferm}
  \end{table}

\section{Conclusions}\label{con}

In this work, we introduced a framework for the construction of exactly solvable Hamiltonians. Our framework has two parts: (1) a polynomial sized compact Lie algebra $\mathcal{A}$, together with the associated Lie group action, (2) the existence of an abelian symmetry group $\mathcal{G}$ of $\mathcal{A}$. The exactly solvable Hamiltonians are formed as linear combinations of the Lie algebra generators, but where the coefficients are polynomials of the symmetry group generators, which are elements of the group algebra $\mathbb{R}[\mathcal{G}]$. The group algebra elements can serve as ``generalized coefficients'', in the sense that any assignment of eigenvalues to all of the group generators can be substituted in to transform the coefficients into real numbers. Our framework goes beyond standard Lie algebra solvability via the maximal torus theorem since the resulting Hamiltonians are no longer elements of the Lie algebra. 

In principle, our framework can be used to construct exactly solvable Hamiltonians in any Lie algebra relevant to a quantum many-body problem. In the qubit algebra, application of this framework to the $so(L + 1)$ algebra generated by $L$ mutually anti-commuting Pauli operators yields non-contextual Pauli Hamiltonians. In the fermionic algebra, mean-field solvable Hamiltonians on $N$ fermionic modes can be expressed in terms of a mean-field $u(N-K)$ algebra on $N-K$ modes, and symmetry projectors on the remaining $K$ modes. These Hamiltonians can be applied to find compact decompositions of any Hamiltonian into exactly solvable fragments, and in the qubit case, it was shown that usage of non-contextual Pauli Hamiltonians provides a reduction in the total number of fragments.

It is known from previous work\cite{yenCartanSubalgebraApproach2021} that a reduction in the number of measurable fragments is not sufficient for a reduction in the overall measurement cost, as reducing the fragment count may yield an increase in the resultant fragments variances, which is the relevant parameter determining the cost of computing the energy functional via measurements. Thus, to fully determine the potential of non-contextual Pauli Hamiltonians as measurable fragments of a general Hamiltonian, a more in depth analysis of variance minimization techniques\cite{Choi2022ghostCS} applied to non-contextual Pauli Hamiltonians needs to be done. For fermionic Hamiltonians, construction of Hamiltonians derived from partitioning of spin orbitals into disjoint subsets, as was done in this work, opens the possibility of new exactly solvable Hamiltonians which go beyond mean-field solvable, by allowing for the possibility of correlations within the subsets.

\begin{acknowledgements}
S.P., T.Y., and A.F.I. are grateful to the Natural Sciences and Engineering Research Council of Canada for financial support. This research was enabled in part by support provided by Compute Ontario and the Digital Research Alliance of Canada. 
\end{acknowledgements}

\appendix

\section{Diagonalization of Block Hamiltonians}
\label{appendix_a}

Exact solvability of the Hamiltonians presented in this work relies on the following intermediate result, stated more generally in terms of linear operators on any finite dimensional vector space. Let $\hat{U}_1=e^{\hat{X}_1},\ldots,\hat{U}_N=e^{\hat{X}_N}$ denote a collection of unitaries with anti-Hermitian generators $\hat{X}_1,\ldots,\hat{X}_N$. Let $\hat{P}_1,\ldots,\hat{P}_N$ denote a set of projectors such that $\hat{P}_i \hat{P}_j = \delta_{ij}\hat{P}_j$. If $[\hat{U}_i, \hat{P}_j] = 0$ for all $i,j$, then
\begin{equation}
    e^{\sum_{i=1}^{N} \hat{X}_i \hat{P}_i} = \sum_{i=1}^{N} \hat{U}_i \hat{P}_i + \left(1 - \sum_{i=1}^{N} \hat{P}_i\right).
\end{equation}
The proof is by strong induction on $N$. For $N = 1$, using the Taylor expansion together with $[\hat{U}_1, \hat{P}_1] = 0$ yields
\begin{align}
    e^{\hat{X}_1 \hat{P}_1} &= \sum_{k=0}^{\infty} \frac{(\hat{X}_1 \hat{P}_1)^k}{k!}\nonumber\\
    &= 1 + \left(\sum_{k=1}^{\infty}\frac{\hat{X}_1^k}{k!}\right)\hat{P}_1\nonumber\\
    &= 1 + (e^{\hat{X}_1} - 1)\hat{P}_1\nonumber\\
    &= \hat{U}_1\hat{P}_1 + (1 - \hat{P}_1)\label{basecase}.
\end{align}
Now, assuming the result holds for all $k < N$, using $[\hat{X}_i \hat{P}_i, \hat{X}_N \hat{P}_N] = 0$ for all $i$ allows us to separate the exponent and apply the induction hypothesis to get
\begin{align}
    e^{\sum_{i=1}^{N} \hat{X}_i \hat{P}_i} &= e^{\sum_{i=1}^{N-1} \hat{X}_i \hat{P}_i}e^{\hat{X}_N \hat{P}_N}\nonumber\\
    &= \left[\sum_{i=1}^{N-1} \hat{U}_i \hat{P}_i + \left(1 - \sum_{i=1}^{N - 1} \hat{P}_i\right)\right]\nonumber\\ 
    &\times \Big(\hat{U}_N \hat{P}_N + (1 - \hat{P}_N)\Big).
\end{align}
Then, repeated application of the commutativity of $\hat{U}$'s and $\hat{P}$'s together with $\hat{P}_i \hat{P}_N = \delta_{iN} \hat{P}_N = 0$ after distributing the product will yield the desired result. Next, consider the Hamiltonian $\hat{H} = \sum_{i=1}^{N} \hat{H}_i \hat{P}_i$, where for all $i,j, [\hat{H}_i, \hat{P}_j] = 0$, and the $\hat{P}_j$ form a complete orthonormal projector set. Suppose furthermore that each $\hat{H}_i$ is diagonalized by the unitary $\hat{U}_i = e^{\hat{X}_i}$:
\begin{equation}
    \hat{H}_i = \hat{U}_i^\dagger \hat{D}_i \hat{U}_i.
\end{equation}
Then, the unitary which diagonalizes $\hat{H}$ can be written as an exponential in the following way
\begin{equation}
    \hat{U}_d = e^{\sum_{i=1}^{N} \hat{X}_i \hat{P}_i}.
\end{equation}
To prove this, first, write $\hat{U}_d = \sum_{i=1}^{N} \hat{U}_i \hat{P}_i$. Then
\begin{align}
    \hat{U}_d \hat{H} \hat{U}_d^\dagger &= \sum_{i,j,k=1}^{N} \hat{U}_i \hat{P}_i \hat{H}_j \hat{P}_j \hat{P}_k^\dagger \hat{U}_k^\dagger  \nonumber\\
    &= \sum_{i=1}^{N} \hat{D}_i \hat{P}_i.
\end{align}

It is often the case that the unitary of interest is presented as a product of unitaries whose generators are known, rather than as the exponential of a single generator. These results can be extended to unitaries which are presented in a product form in the following way. Consider a unitary $\hat{U} = e^{\hat{X}}$ which admits a product decomposition of the following form
\begin{equation}
    \hat{U} = \prod_{i=1}^{N} e^{\hat{X}_i},
\end{equation}
and suppose $\hat{P}$ is a projector which commutes with all of the $\hat{X}_i$. Then
\begin{equation}
    \hat{U}\hat{P} + (1 - \hat{P}) = \prod_{i=1}^{N} e^{\hat{X}_i \hat{P}}.
\end{equation}
To prove this, write
\begin{equation}
    \prod_{i=1}^{N} e^{\hat{X}_i \hat{P}} = \prod_{i=1}^{N} \big(e^{\hat{X}_i} \hat{P} + (1 - \hat{P})\big).
\end{equation}
When expanding the above product, all of the cross terms will have a factor $\hat{P}(1-\hat{P}) = 0$, implying that only the first and last terms survive. Thus, we get
\begin{align}
    \prod_{i=1}^{N} e^{\hat{X}_i \hat{P}} &= \prod_{i=1}^{N} e^{\hat{X}_i} \hat{P}^N + (1 - \hat{P})^N\nonumber\\
    &= \hat{U}\hat{P} + (1 - \hat{P}).
\end{align}
With this result, given unitaries $\hat{U}_i = e^{\hat{X}_i}$, each with product decomposition $\hat{U}_i = \prod_{j=1}^{N_i} e^{\hat{X}_{ij}},$ and a complete orthonormal set of projectors $\hat{P}_k$ with $[\hat{X}_{ij}, \hat{P}_k] = 0$ for all $i,j,k$, we can obtain a product decomposition of $\sum_{i=1}^{N} \hat{U}_i \hat{P}_i$ as follows
\begin{align}
    \sum_{i=1}^{N} \hat{U}_i \hat{P}_i &= e^{\sum_{i=1}^{N} \hat{X}_i \hat{P}_i}\nonumber\\
    &= \prod_{i=1}^{N} e^{\hat{X}_i \hat{P}_i}\nonumber\\
    &= \prod_{i=1}^{N} \prod_{j=1}^{N_i} e^{\hat{X}_{ij}\hat{P}_j}. \label{prod_form}
\end{align}

\section{Factorization of Non-Contextual Pauli Hamiltonian}
\label{appendix_b}

Starting from the non-contextual set $\mathcal{S}$, the following procedure, summarized from \citet{Kirby2020_contextual_simulation} allows you to write the non-contextual Pauli Hamiltonian $\hat{H}_\text{nc}$ whose Pauli terms come from $\mathcal{S}$ in the form of \eq{qubit_nc}. Denote $\mathcal{Z} = \{\hat{Z}_\alpha : 1 \leq \alpha \leq |\mathcal{Z}|\}$, and $\mathcal{C}_i = \{\hat{A}_{ij} : 1 \leq j \leq |\mathcal{C}_i|\}$. Then, the non-contextual Pauli Hamiltonian has the following form
\begin{equation}
    \hat{H}_{\text{nc}} = \sum_{\alpha=1}^{|\mathcal{Z}|} h_\alpha \hat{Z}_\alpha + \sum_{i=1}^{L} \sum_{j=1}^{|\mathcal{C}_i|} h_{ij} \hat{A}_{ij} \label{generic_nc}.
\end{equation}
We can obtain a mutually anti-commuting set $\{\hat{A}_i\}_{i=1}^{L}$ of Pauli operators by choosing a single representative of each commutation class, which, without loss of generality, we take to be $\hat{A}_i = \hat{A}_{i1}$. Then, defining $\hat{C}_{ij} = \hat{A}_{ij} \hat{A}_{i}$, so that $\hat{A}_{ij} = \hat{C}_{ij} \hat{A}_{i}$ from the involutory property of Pauli operators, we have
\begin{equation}
    \hat{H}_{\text{nc}} = \sum_{\alpha=1}^{|Z|} h_\alpha \hat{Z}_\alpha + \sum_{i=1}^{L} \left(\sum_{j=1}^{|C_i|}h_{ij} \hat{C}_{ij}\right)\hat{A}_{i}. \label{nc_form2}
\end{equation}
The set $\mathcal{Z} \cup \{\hat{C}_{ij}\}$ mutually commute, and commute with everything in $\mathcal{S}$, implying that $\mathcal{Z} \cup \{\hat{C}_{ij}\}$ generates some finite abelian group $\mathcal{G}$ with minimal generating set $G = \{\hat{C}_1,\ldots,\hat{C}_K\}$. Thus, we can write $\hat{H}_{\text{nc}}$ in terms of the mutually anti-commuting operators $\hat{A}_{i}$ and the group generators $\hat{C}_k$ since we can write linear combinations of elements of $\mathcal{Z} \cup \{\hat{C}_{ij}\}$ as polynomials $p_0, p_i$ over $G$:
\begin{align}
    p_0(\hat{C}_1,\ldots,\hat{C}_k) &= \sum_{\alpha=1}^{|Z|} h_\alpha \hat{Z}_\alpha\nonumber\\
    p_i(\hat{C}_1,\ldots,\hat{C}_k) &= \sum_{j=1}^{|C_i|}h_{ij} \hat{C}_{ij},\label{nc_form4}
\end{align}
from which it follows that
\begin{equation}
    \hat{H}_{\text{nc}} = \sum_{i=1}^{L} p_i(\hat{C}_1,\ldots,\hat{C}_k) \hat{A}_i + p_0(\hat{C}_1,\ldots,\hat{C}_k).
\end{equation}
Thus, the non-contextual Hamiltonian in \eq{generic_nc} is equal to an exactly solvable linear combination of anti-commuting Pauli operators, where the coefficients of the linear combination have been replaced by fully commuting operators. Combining this result with the discussion in Sec. \ref{app_q} demonstrates that this property completely characterizes the non-contextual Pauli Hamiltonians.

\section{Details of Electronic Hamiltonians}\label{app_comp}

All electronic Hamiltonians used in this work were generated in the STO-3G basis and using the Bravyi-Kitaev transformation, as implemented in the Openfermion\cite{mccleanOpenFermionElectronicStructure2020} package using PySCF.\cite{sunPySCFPythonbasedSimulations2018,sunRecentDevelopmentsPySCF2020} The nuclear geometries used were:
\begin{itemize}
    \item H$_2$: R$(\rm H-H) = 1 \AA$ 
    \item H$_4$: R$(\rm H-H) = 1 \AA$ with colinear atomic arrangement
    \item LiH:  R$(\rm Li-H) = 1 \AA$ 
    \item BeH$_2$:  R$(\rm Be-H) = 1 \AA$ with colinear atomic arrangement
    \item H$_2$O: R$(\rm O-H) = 1 \AA$ with $\angle \rm HOH = 107.6^\circ$ 
    \item NH$_3$: R$(\rm N-H) = 1 \AA$ with $\angle \rm HNH = 107^\circ$ 
\end{itemize}

\bibliography{library}
\end{document}